\documentclass[aps,prd,a4paper,preprintnumbers,amsmath,amssymb,superscriptaddress,twocolumn,floatfix,showpacs,nofootinbib]{revtex4}

\usepackage{amssymb}

\usepackage{float,graphicx}

\begin{document}

\title{The Cosmology of Ricci-Tensor-Squared Gravity in the Palatini
Variational Approach}

\author{Baojiu~Li }
\email[Email address: ]{b.li@damtp.cam.ac.uk}
\affiliation{Department of Applied Mathematics and Theoretical
Physics, Centre for Mathematical Sciences, Wilberforce Road,
University of Cambridge, Cambridge CB3 0WA, United Kingdom}

\author{John~D.~Barrow}
\email[Email address: ]{j.d.barrow@damtp.cam.ac.uk}
\affiliation{Department of Applied Mathematics and Theoretical
Physics, Centre for Mathematical Sciences, Wilberforce Road,
University of Cambridge, Cambridge CB3 0WA, United Kingdom}

\author{David~F.~Mota }
\email[Email address: ]{d.mota@thphys.uni-heidelberg.de}
\affiliation{Institute of Theoretical Physics, University of
Heidelberg, 69120 Heidelberg, Germany}

\date{\today}

\begin{abstract}

We consider the cosmology of the Ricci-tensor-squared gravity in
the Palatini variational approach. The gravitational action of
standard general relativity is modified by adding a function
$f(R^{ab}R_{ab})$ to the Einstein-Hilbert action, and the Palatini
variation is used to derive the field equations. A general method
of obtaining the background and first-order covariant and
gauge-invariant perturbation equations is outlined. As an example,
we consider the cosmological constraints on such theories arising
from the supernova type Ia and cosmic microwave background
observations. We find that the best fit to the data is a non-null
leading-order correction to Einstein gravity, but the current data
exhibit no significant preference over the concordance model. The
growth of non-relativistic matter density perturbations at late
times is also analyzed, and we find that a scale-dependent
(positive or negative) sound-speed-squared term generally appears
in the growth equation for small-scale density perturbations. We
also estimate the observational bound imposed by the matter power
spectrum for the model with $f(\mathbf{R}^{ab}\mathbf{R}_{ab}) =
\alpha(\mathbf{R}^{ab}\mathbf{R}_{ab})^{\beta }$ to be roughly
$|\beta| \lesssim \mathcal{O}(10^{-5})$ so long as the dark matter
does not possess compensating anisotropic stresses.

\end{abstract}


\maketitle

\section{Introduction}

\label{sect:Introduction}

The accumulating astronomical evidence for an accelerating cosmic
expansion has stimulated many investigations into the nature of
the dark energy, or other possible deviant gravitational effects,
which might be responsible for this unexpected dynamics (for a
review see, \emph{e.g.}, \cite{DEReview}). Besides proposing to
add some new (and purely theoretical) matter species into the
energy budget of the universe, many investigators have also
focused their attentions on modifying general relativity (GR) on
the largest scales, so as to introduce significant modifications
in the behaviour of gravity at late times when it is comparatively
weak. One example of the latter sort is provided by the family of
$f(R)$ gravity models, which had also been considered before the
discovery of cosmic acceleration (see for example
Refs.~\cite{BOtt, BCot, Maeda1989}) with reference to alternative
forms of inflation and the existence of singularities. In
Refs.~\cite{Carroll2005, Easson2005}, the authors discuss a
specific model where the correction to GR is a polynomial function
of the $R^{2},R^{ab}R_{ab}$ and $R^{abcd}R_{abcd}$ quadratic
curvature invariants (here, $R,R_{ab}$ and $R_{abcd}$ are
respectively the Ricci scalar, Ricci tensor and Riemann tensor
calculated in the standard way from the physical metric $g_{ab}$)
and showed that there exist late-time accelerating attractors in
Friedmann cosmological solutions to the theory. Barrow and Clifton
established the general existence conditions for de Sitter,
Einstein static, G\"{o}del universes in theories where the
Lagrangian is an arbitrary function of these three invariants
\cite{Clifton2005}.

When the Ricci scalar $R$ in the Einstein-Hilbert action is
replaced by some general functions of $R$ and $R^{ab}R_{ab}$, it
becomes necessary to distinguish between two different variational
approaches to deriving the field equations. In the metric
approach, as in Refs.~\cite{Carroll2005, Easson2005}, the metric
components $g_{ab}$ are the only variational quantities and the
field equations are generally of fourth-order, which makes the
theories phenomenologically richer but more stringently
constrained in most cases. Within the Palatini variational
approach, on the other hand, we treat the metric $g_{ab}$ and the
connection $\Gamma _{bc}^{a}$ as independent variables and
extremize the action with respect to both of them; the resulting
field equations are second order and easier to solve. The Palatini
$f(R)$ gravity is also proposed as an alternative to dark energy
in a series of works \cite{Vollick2003, Allemandi2004a,
Allemandi2005}. There has since been growing interest in these
modified gravity theories: for the local tests of the Palatini and
metric $f(R)$ gravity models see \cite{PalatiniLGT, MetricLGT};
for the cosmologies of these two classes of models see
\cite{PalatiniCT, MetricCT, Koivisto2006, Li2006a, Li2006b}.

Both approaches to modifying gravity are far from problem-free. In
the metric $f(R)$ gravity models, the theory is conformally
related to standard GR plus a self-interacting scalar field
\cite{BCot}, which generally introduces extra forces inconsistent
with solar system tests \cite{MetricLGT}. The Palatini approach,
on the other hand, generally leads to a large (or even negative)
sound-speed-squared term in the growth equation of the
non-relativistic matter perturbations on small scales. This
induces effects on the cosmic microwave background (CMB) and the
matter clustering power-spectra which deviate unacceptably from
those which are observed \cite{Koivisto2006, Li2006a, Li2006b}.
Again, these examples highlight the difficulties encountered when
trying to make modifications to standard GR which are compatible
with observations.

In this work we will focus on the Ricci-squared gravity models
within the Palatini variational approach, which we also denote by
the $f(\mathbf{R}^{ab}\mathbf{R}_{ab})$ gravity. It turns out that
the Ricci tensor, $\mathbf{R}_{ab},$ and Ricci scalar,
$\mathbf{R}$, appearing in the field equations in the Palatini
approach are not the ones calculated from the physical metric,
$g_{ab}$, (we consider the metric $g_{ab}$ as the physical one
because it is this metric which the matter Lagrangian density
$\mathcal{L}_{m}$ depends on and the energy-momentum conservation
law holds with respect to) as in GR, and we denote the GR
equivalents by $R_{ab}$ and $R$ respectively to distinguish them
from the Palatini quantities). Such a modification of gravity has
indeed been considered in \cite{Allemandi2004b} and shown to give
an accelerating cosmology. However, our work differs from
\cite{Allemandi2004b} in that we replace the Ricci scalar in the
gravitational action with
$\mathbf{R}+f(\mathbf{R}^{ab}\mathbf{R}_{ab})$ rather than simply
$f(\mathbf{R}^{ab}\mathbf{R}_{ab}),$ and we concentrate more on
the cosmology at the first-order perturbation level, especially
the late-time cold-dark-matter (CDM) density perturbation growth.
We emphasize the similarity to the Palatini $f(\mathbf{R})$
gravity models also.

Our presentation is organized as follows. In
Sec.~\ref{sect:Equations} we briefly introduce the model and
outline the methods used to derive the background and first-order
covariant and gauge-invariant (CGI) perturbed field equations. In
Sec.~\ref{sect:Background}, we present the modified Friedmann
equation and apply it to a specific family of theories with
$f(\mathbf{R}^{ab}\mathbf{R}_{ab})=\alpha
(\mathbf{R}^{ab}\mathbf{R}_{ab})^{\beta }$; the constraints on the
parameter space $(\beta ,\Omega_{m})$ from cosmological data are
also given. Then, in Sec.~\ref{sect:Perturbations}, we analyse the
growth of CDM density perturbations at late cosmological times for
this model. Since this analysis shares some similarities with the
Palatini $f(\mathbf{R})$ gravity models, we also present a similar
discussion of the latter for comparison. Our discussion and
conclusions are presented in Sec.~\ref{sect:Conclusions}.

Throughout this work our convention is chosen as
$[\nabla_{a},\nabla_{b}]u^{c}=R_{ab\ d}^{\ \
c}u^{d},R_{ab}=R_{acb}^{\ \ \ c}$ where $ a,b,\cdots $ run over
$0,1,2,3$ and $c=\hslash =1$; the metric signature is $(+,-,-,-)$
and the universe is assumed to be spatially flat and filled with
CDM and black body radiation.

\section{Field Equations in $f(\mathbf{R}^{ab}\mathbf{R}_{ab})$ Gravity}

\label{sect:Equations}

In this section we briefly introduce the main ingredients of
$f(\mathbf{R}^{ab}\mathbf{R}_{ab})$ gravity and outline the
strategy for deriving the general perturbation equations that
govern the dynamics of small inhomogeneities in the cosmological
models that arise in this theory.

\subsection{The $f(\mathbf{R}^{ab}\mathbf{R}_{ab})$ Gravity Model}

We will start our discussion with the modified Einstein-Hilbert
action in the present model,
\begin{eqnarray}  \label{eq:1}
S &=& \int d^{4}x\sqrt{-g}\left[
\frac{\mathbf{R}+f(\mathbf{R}^{ab}\mathbf{R}_{ab})}{2\kappa}+\mathcal{L}_{m}\right],
\end{eqnarray}
in which $\kappa =8\pi G_{\mathrm{N}},$ with $G_{\mathrm{N}}$ the
Newtonian gravitational constant. Here,
$\mathbf{R}_{ab}=\mathbf{R}_{ab}(\mathbf{\Gamma }_{bc}^{a})$ is
given by
\begin{eqnarray}  \label{eq:2}
\mathbf{R}_{ab} &=&
\mathbf{\Gamma}_{ab,c}^{c}-\mathbf{\Gamma}_{ac,b}^{c}+\mathbf{\Gamma}_{cd}^{c}\mathbf{\Gamma}_{ab}^{d}-\mathbf{\Gamma}_{ad}^{c}
\mathbf{\Gamma }_{cb}^{d}
\end{eqnarray}
and $\mathbf{R}=g^{ab}\mathbf{R}_{ab}$; note that
$\mathbf{\Gamma}_{bc}^{a}$ is a new and independent variable with
respect to which we extremise the action, and is different from
the Christoffel symbol $\Gamma _{bc}^{a}$ calculated using the
metric $g_{ab}$. $\mathbf{R}_{ab}$ is assumed to be a symmetric
tensor (if it contains an antisymmetric part then the field
equation will be spoiled as discussed in \cite{Borowiec1998}) and
$g_{ab}$ could be used to raise or lower its indices. Varying the
action Eq.~(\ref{eq:1}) with respect to the metric $g_{ab}$ (note
that $\delta\mathbf{\Gamma}/\delta g = 0$ as they are independent)
gives the modified Einstein equations:
\begin{eqnarray}  \label{eq:3}
\mathbf{R}_{ab}+2F\mathbf{R}_{a}^{c}\mathbf{R}_{bc}-\frac{1}{2}g_{ab}\left[\mathbf{R}+f(\mathbf{R}^{ab}\mathbf{R}_{ab})\right]
&=& \kappa T_{ab}^{f}\ \
\end{eqnarray}
where $F=\partial f/\partial \mathbf{S}$ with
$\mathbf{S}=\mathbf{R}^{ab} \mathbf{R}_{ab}$ and $T_{ab}^{f}$ is
the energy-momentum tensor of the fluid matter (CDM and
radiation).

On the other hand, varying the action with respect to the new
variable $\mathbf{\Gamma }_{bc}^{a}$ with the relation $\delta
\mathbf{R}_{ab} = \mathbf{D}_{c}(\delta\mathbf{\Gamma}^{c}_{ab}) -
\mathbf{D}_{b}(\delta\mathbf{\Gamma}^{c}_{ca})$, one arrives at
another field equation
\begin{eqnarray}  \label{eq:4}
\mathbf{D}_{e}\left[ \sqrt{\det g}\left(
g^{ab}+2Fg^{ac}\mathbf{R}_{cd}g^{bd}\right) \right] &=& 0,
\end{eqnarray}
where $\mathbf{D}_{a}$ represents the covariant derivative
compatible to $ \mathbf{\Gamma }_{bc}^{a}$ (the covariant
derivative compatible to $g_{ab}$ is denoted, as conventionally,
by $\nabla _{a}$). Just like in the Palatini $ f(\mathbf{R})$
models, this equation implies some relation between the physical
metric $g_{ab}$ and the metric $\mathbf{g}_{ab}$ whose Christoffel
symbol is $\mathbf{\Gamma }_{bc}^{a}$. However, because of the
presence of the second term in the parentheses this relation is
nontrivial and some further algebra will be needed to explicate
it. Before doing that, we will present some preliminary
definitions and expressions, one of which is the notation of $3+1$
decomposition.

\subsection{The $3+1$ Decomposition}

The main idea of $3+1$ decomposition \cite{Ellis1989, Ellis1998,
Challinor1999, Tsagas2007} is to make spacetime splits of physical
quantities with respect to the 4-velocity $u^{a}$ of an observer.
The projection tensor $h_{ab}$ is defined as
$h_{ab}=g_{ab}-u_{a}u_{b}$ and can be used to obtain covariant
tensors perpendicular to $u^{a}$. For example, the covariant
spatial derivative $\hat{\nabla}$ of a tensor field $T_{d\cdot
\cdot \cdot e}^{b\cdot \cdot \cdot c}$ is defined as
\begin{eqnarray}  \label{eq:5}
\hat{\nabla}^{a}T_{d\cdot \cdot \cdot e}^{b\cdot \cdot \cdot c}
&\equiv& h_{i}^{a}h_{j}^{b}\cdot \cdot \cdot \
h_{k}^{c}h_{d}^{r}\cdot \cdot \cdot \ h_{e}^{s}\nabla
^{i}T_{r\cdot \cdot \cdot s}^{j\cdot \cdot \cdot k}.
\end{eqnarray}

The energy-momentum tensor and covariant derivative of the
4-velocity are decomposed respectively as
\begin{eqnarray}  \label{eq:6}
T_{ab} &=& \pi _{ab}+2q_{(a}u_{b)}+\rho u_{a}u_{b}-ph_{ab}, \\
\label{eq:7} \nabla _{a}u_{b} &=& \sigma _{ab}+\varpi
_{ab}+\frac{1}{3}\theta h_{ab}+u_{a}A_{b}.
\end{eqnarray}
In the above, $\pi _{ab}$ is the projected symmetric trace-free
(PSTF) anisotropic stress, $q_{a}$ the heat flux vector, $p$ the
isotropic pressure, $\sigma _{ab}$ the PSTF shear tensor,
$\varpi_{ab}=\hat{\nabla}_{[a}u_{b]}$, is the vorticity, $\theta
=\nabla ^{c}u_{c}\equiv 3\dot{a}/a$ ($a$ is defined here as the
mean expansion scale factor) the volume expansion rate scalar, and
$A_{b}=\dot{u}_{b}$ is the fluid acceleration; the overdots denote
time derivatives expressed as $\dot{\phi}=u^{a}\nabla _{a}\phi$,
brackets mean antisymmetrisation, and parentheses symmetrization.
The velocity normalization is chosen to be $u^{a}u_{a}=1$. The
quantities $\pi_{ab},q_{a},\rho ,p$ are referred to as
\emph{dynamical} quantities and $\sigma _{ab},\varpi _{ab},\theta
,A_{a}$ as \emph{kinematical} quantities. Note that the dynamical
quantities can be obtained from the energy-momentum tensor
$T_{ab}$ through the relations
\begin{eqnarray}  \label{eq:8}
\rho &=& T_{ab}u^{a}u^{b},  \nonumber \\
p &=& -\frac{1}{3}h^{ab}T_{ab},  \nonumber \\
q_{a} &=& h_{a}^{d}u^{c}T_{cd},  \nonumber \\
\pi _{ab} &=& h_{a}^{c}h_{b}^{d}T_{cd}+ph_{ab}.
\end{eqnarray}

Decomposing the Riemann tensor and making use the Einstein
equations, we could obtain, after linearization, the perturbed
(constraint and propagation) equations \cite{Ellis1989, Ellis1998,
Challinor1999, Tsagas2007}. Here, we shall not list all of them
because most are irrelevant for the following discussion; rather
we will use the linearised Raychaudhuri equation
\begin{eqnarray}  \label{eq:9}
\dot{\theta}+\frac{1}{3}\theta
^{2}-\hat{\nabla}^{a}A_{a}+\frac{\kappa }{2}(\rho +3p) &=& 0,
\end{eqnarray}
the linearised conservation equations for the energy density:
\begin{eqnarray}  \label{eq:10}
\dot{\rho}+(\rho +p)\theta +\hat{\nabla}^{a}q_{a} &=& 0,
\end{eqnarray}
and the linearised Friedmann equation
\begin{eqnarray}  \label{eq:11}
\frac{1}{3}\theta ^{2} &=& \kappa\rho .
\end{eqnarray}

The above equations are derived and presented for standard general
relativity, and so the $\rho, p, q_{a}$ variables describe
imperfect fluid matter. For general modified gravity theories,
such as those presented here, the modification to GR might be
parameterized as an effective energy-momentum tensor. In this case
the formalism of these equations is preserved and one just needs
to replace $\rho ,p,q_{a}$ by the total effective quantities of
the same sort: $\rho ^{\mathrm{tot}}, p^{\mathrm{tot}},
q_{a}^{\mathrm{tot}}$ \cite{Hwang1990}.

\subsection{The Field Equations in $f(\mathbf{R}^{ab}\mathbf{R}_{ab})$
Gravity}

In Eq.~(\ref{eq:4}), we see that $\sqrt{\det g}\left(
g^{ab}+2Fg^{ac}\mathbf{R}_{cd}g^{bd}\right) $ is a symmetric
$(2,0)$ tensor density of weight 1, and so we can introduce a new
metric $\mathbf{g}_{ab}$ by means of the following relation
\begin{eqnarray}  \label{eq:12}
\sqrt{\det \mathbf{g}}\mathbf{g}^{ab} &=& \sqrt{\det g}\left(
g^{ab}+2Fg^{ac}\mathbf{R}_{cd}g^{bd}\right) ,
\end{eqnarray}
where the Levi-Civita connection of the metric $\mathbf{g}_{ab}$
is just $\mathbf{\Gamma }_{bc}^{a}$, as we referred to above.

To go further, we need to express $\mathbf{R}_{ab}$ explicitly.
This is easy to do in principle, because Eq.~(\ref{eq:3}) is just
an algebraic equation for $\mathbf{R}_{ab}$. To see this, let us
write the symmetric tensor $\mathbf{R}_{ab}$ in a general way as
\begin{eqnarray}  \label{eq:13}
\mathbf{R}_{ab} &=& \Delta u_{a}u_{b}+\Xi
h_{ab}+2u_{(a}\Upsilon_{b)}+\Sigma _{ab}
\end{eqnarray}
where $u_{a}$ is the 4-velocity of the observer referred to above.
Substituting Eqs.~(\ref{eq:6}, \ref{eq:13}) into Eq.~(\ref{eq:3}),
we get
\begin{eqnarray}
&&\Delta u_{a}u_{b}+\Xi h_{ab}+2u_{(a}\Upsilon _{b)}+\Sigma _{ab}
\nonumber\\
&&+2F\left[ \Delta ^{2}u_{a}u_{b}+\Xi ^{2}h_{ab}+2(\Delta +\Xi
)u_{(a}\Upsilon _{b)}+2\Xi \Sigma _{ab}\right]  \nonumber \\
&&-\frac{1}{2}\left( \Delta +3\Xi +f\right)
u_{a}u_{b}-\frac{1}{2}\left(
\Delta +3\Xi +f\right) h_{ab}  \nonumber \\
&=&\kappa \left(\rho^{f}u_{a}u_{b}-p^{f}h_{ab} + 2u_{(a}q_{b)}^{f}
+ \pi_{ab}^{f}\right),  \nonumber
\end{eqnarray}
which leads to the following four equations:
\begin{eqnarray}  \label{eq:14}
\Delta +2F\Delta ^{2}-\frac{1}{2}\left( \Delta +3\Xi +f\right) &=&
\kappa\rho ^{f}, \\
\label{eq:15} \Xi +2F\Xi ^{2}-\frac{1}{2}\left( \Delta +3\Xi
+f\right) &=& -\kappa p^{f},\\
\label{eq:16} \left[ 1+2F(\Delta +\Xi )\right] \Upsilon _{a} &=& \kappa q_{a}^{f}, \\
\label{eq:17} (1+4F\Xi )\Sigma _{ab} &=& \kappa \pi _{ab}^{f},
\end{eqnarray}
where $f,F$ are functions of
$\mathbf{R}^{ab}\mathbf{R}_{ab}=\Delta^{2}+3\Xi ^{2}$. Thus given
the specified form of $f,$ and the values of $\rho ^{f}, p^{f}$,
the quantities $\Delta, \Xi$ can be obtained from
Eqs.~(\ref{eq:14}, \ref{eq:15}), at least numerically. Then,
$\Upsilon_{a}$ and $\Sigma _{b}$ can also be calculated from
Eqs.~(\ref{eq:16}, \ref{eq:17}) provided the values of $q_{a}^{f}$
and $\pi_{ab}^{f}$ are given. Note that $\Upsilon_{a}$ and
$\Sigma_ {ab}$ are nonzero only at first order in perturbation.
Taking the time derivatives of Eqs.~(\ref{eq:14}, \ref{eq:15}),
and using the background values of $\Delta, \Xi$, we could easily
obtain $\dot{\Delta}$ and $\dot{\Xi}$ by solving the two linear
algebraic equations. Similarly, $\hat{\nabla}_{a}\Delta $ and
$\hat{\nabla}_{a}\Xi $ could be worked out (here, $\hat{\nabla}$
is the spatial derivative). In what follows, we shall assume that
$\Delta ,\Xi $ and their derivatives have been calculated.

The next step is to find out the relation between $g_{ab}$ and
$\mathbf{g}_{ab}$. We could rewrite Eq.~(\ref{eq:12}) as
\begin{eqnarray}  \label{eq:18}
\sqrt{\det \mathbf{g}}\mathbf{g}^{ab} &=& \sqrt{\det
g}g^{ac}(\delta_{c}^{b}+2F\mathbf{R}_{c}^{b}).
\end{eqnarray}
Taking the determinants of both sides and equating we get
\begin{eqnarray}  \label{eq:19}
\det\mathbf{g} &=& \det g\cdot\det \mathbf{P,}
\end{eqnarray}
with
\begin{eqnarray}  \label{eq:20}
\mathbf{P}_{b}^{a} &=& \delta _{b}^{a}+2F\mathbf{R}_{b}^{a}.
\end{eqnarray}

Thus, we conclude from Eq.~(\ref{eq:18}) that
\begin{eqnarray}  \label{eq:21}
\mathbf{g}^{ab} &=& \frac{\sqrt{\det g}}{\sqrt{\det
\mathbf{g}}}g^{ac}\mathbf{P}_{c}^{b}  \nonumber \\
&=&\frac{1}{\sqrt{\det \mathbf{P}}}\left( g^{ab}+2F\mathbf{R}^{ab}\right); \\
\label{eq:22} \mathbf{g}_{ab} &=& \frac{\sqrt{\det
\mathbf{g}}}{\sqrt{\det g}}g_{ac}\left(\mathbf{P}^{-1}\right) _{b}^{c}  \nonumber \\
&=& \sqrt{\det \mathbf{P}}\left(\mathbf{P}^{-1}\right)_{ab}.
\end{eqnarray}
Obviously, $\det \mathbf{P}$ and $\mathbf{P}^{-1}$ need to be
evaluated respectively. For $\det \mathbf{P}$, we have
\begin{eqnarray}
&&\det \mathbf{P}  \nonumber \\
&=& \det[(1+2F\Delta )u^{a}u_{b}+(1+2F\Xi)h_{b}^{a}\nonumber\\
&& +4Fu^{(a}\Upsilon _{b)}+2F\Sigma _{b}^{a}]\nonumber.
\end{eqnarray}
To calculate this, let us write $g_{00} = a^{2}(1+2\Psi),
g_{0\alpha} = g_{\alpha0} = a^{2}B_{\alpha}, g_{\alpha\beta} =
a^{2}\left[(1+2H_{L})\gamma_{\alpha\beta} +
2H_{T\alpha\beta}\right]$ and $u_{0} = a(1+\Psi), u_{\alpha} =
-a(v_{\alpha}-B_{\alpha})$ where $\alpha, \beta$ run over $1, 2,
3$, $\Psi, B_{\alpha}, H_{L}$ and $H_{T\alpha\beta}$ are first
order metric variables of which $H_{T\alpha\beta}$ is traceless,
$v_{\alpha}$ is the spatial component of $u_{a}$, and
$\gamma_{\alpha\beta}$ is the metric of 3 dimensional flat space.
As a result $g^{00} = a^{-2}(1-2\Psi), g^{0\alpha} = g^{\alpha0} =
-a^{-2}B^{\alpha}, g^{\alpha\beta} =
(1-2H_{L})\gamma^{\alpha\beta} - 2H^{\alpha\beta}_{T}$ and $u^{0}
= a^{-1}(1-\Psi), u^{\alpha} = -a^{-1}v^{\alpha}$. From these
expressions the components of $h^{a}_{b}$ can also be obtained and
one can substitute all these quantities into the above equation to
get $\det\mathbf{P}$. Since $\Upsilon_{a}$ and $\Sigma_{ab}$ are
only of first order and because $\Sigma_{ab}$ is traceless, it is
then not difficult to see that up to first order (note that the
facts $u^{a}\Upsilon_{a} = u_{a}\Upsilon^{a} = 0$ and
$u^{a}\Sigma_{ab} = 0$ indicate that $\Upsilon^{0} = \Upsilon_{0}
= 0$ and $\Sigma^{0}_{a} = \Sigma^{a}_{0} = 0$)
\begin{eqnarray}  \label{eq:23}
\det \mathbf{P} &=& (1+2F\Delta )(1+2F\Xi )^{3}.
\end{eqnarray}

For $(\mathbf{P}^{-1})_{b}^{a}$, we know that it is symmetric as
the inverse matrix of a symmetric matrix, and so could be written
as
\begin{eqnarray}  \label{eq:24}
\left( \mathbf{P}^{-1}\right) _{b}^{a} &=&
Au^{a}u_{b}+Bh_{b}^{a}+2u^{(a}C_{b)}+D_{b}^{a}.
\end{eqnarray}
Using
\begin{eqnarray}
&&\mathbf{P}_{c}^{b}  \nonumber \\
&=& (1+2F\Delta )u^{b}u_{c}+(1+2F\Xi
)h_{c}^{b}+4Fu^{(b}\Upsilon_{c)}+2F\Sigma_{c}^{b}  \nonumber
\end{eqnarray}
and
\begin{eqnarray}
\left(\mathbf{P}^{-1}\right) _{b}^{a}\mathbf{P}_{c}^{b} &=&
\delta_{c}^{a} \nonumber
\end{eqnarray}
it is then easy to obtain, to first order, that
\begin{eqnarray}  \label{eq:25}
A &=& \frac{1}{1+2F\Delta };  \nonumber \\
B &=& \frac{1}{1+2F\Xi };  \nonumber \\
C_{a} &=& -\frac{2F}{(1+2F\Delta )(1+2F\Xi )}\Upsilon _{a};  \nonumber \\
D_{ab} &=& -\frac{2F}{(1+2F\Xi )^{2}}\Sigma _{ab}.
\end{eqnarray}

As a result, we have now the relations between the two metrics
$\mathbf{g}_{ab}$ and $g_{ab}$ and their inverses as
\begin{eqnarray}  \label{eq:26}
\mathbf{g}_{ab} &=&\lambda g_{ab}+\xi _{ab}, \\
\label{eq:27} \mathbf{g}^{ab} &=&\frac{1}{\lambda }g^{ab} +
\zeta^{ab},
\end{eqnarray}
where
\begin{eqnarray}  \label{eq:28}
\lambda &=& \sqrt{(1+2F\Delta)(1+2F\Xi)}, \\
\label{eq:29} \omega &=& \frac{1+2F\Xi }{1+2F\Delta }, \\
\label{eq:30} \xi _{ab} &=& \lambda (\omega -1)u_{a}u_{b}
- 4\sqrt{\omega}Fu_{(a}\Upsilon_{b)} - \frac{2F}{\sqrt{\omega}}\Sigma_{ab}, \\
\label{eq:31} \zeta ^{ab} &=& \frac{1}{\lambda }\left(
\frac{1}{\omega }-1\right) u^{a}u^{b}  +
\frac{1}{\lambda^{2}}\frac{2F}{\sqrt{\omega}}\left[2u^{(a}\Upsilon^{b)}
+ \Sigma^{ab}\right].
\end{eqnarray}

A discussion of how the two Ricci tensors
$\mathbf{R}_{ab}(\mathbf{\Gamma}^{a}_{bc})$ and $R_{ab}(g_{ab})$
are related to one another is given in the appendix, with the help
of which the Einstein equation Eq.~(\ref{eq:3}) can be rewritten
as
\begin{eqnarray}  \label{eq:32}
R_{ab}-\frac{1}{2}g_{ab}R &=& \kappa T_{ab}^{f}+\kappa
T_{ab}^{eff}
\end{eqnarray}
where
\begin{eqnarray}  \label{eq:33}
\kappa T_{ab}^{eff} &\equiv& \frac{1}{2}g_{ab}(f+\delta R)-\delta R_{ab}-2F%
\mathbf{R}_{a}^{c}\mathbf{R}_{cb},
\end{eqnarray}
and (see the appendix for a definition of the tensor
$\gamma_{bc}^{a}$)
\begin{eqnarray}  \label{eq:34}
\delta R_{ab} &=&
\nabla_{c}\gamma_{ab}^{c}-\nabla_{b}\gamma_{ac}^{c} +
\gamma _{ab}^{d}\gamma _{cd}^{c}-\gamma_{ac}^{d}\gamma _{bd}^{c}, \\
\label{eq:35} \delta R &=& g^{ab}\delta R_{ab}.
\end{eqnarray}

With the aid of Eqs.~(\ref{eq:8}, \ref{eq:33}, \ref{eq:34},
\ref{eq:35}) one could identify $\rho ^{eff}, p^{eff},
q_{a}^{eff}$ and $\pi _{ab}^{eff}$ and express them in terms of
$\omega, \lambda, F, \Delta $ and $\Xi,$ which are functions of
$\rho ^{f}$ and $p^{f}$ (c.f., Eqs.~(\ref{eq:14}, \ref{eq:15},
\ref{eq:28}, \ref{eq:29})), and $\Upsilon_{a}, \Sigma_{ab}$ which
are also functions of $q_{a}^{f},\pi _{ab}^{f}$ (c.f.,
Eqs.~(\ref{eq:16}, \ref{eq:17})). However, from Eqs.~(\ref{eq:26}
- \ref{eq:31}, \ref{eq:A6}, \ref{eq:A7}) one can see that this
process will involve a lot of calculation. In the present work we
will not perform detailed numerical calculations of the
perturbation equations of the $f(\mathbf{R}^{ab}\mathbf{R}_{ab})$
model; instead, in the next two sections of the paper we will:

1. Study the background evolutions of general Ricci-squared
gravity models. As an example, we will consider a specific family
of theories with $f(\mathbf{R}^{ab}\mathbf{R}_{ab}) =
\alpha(\mathbf{R}^{ab}\mathbf{R}_{ab})^{\beta }$, and constrain
the allowed $(\alpha ,\beta )$ parameter space using data sets
supernovae (SNe) luminosity distances and the CMB shift parameter.

2. Present a simple argument to show that this class of modified
gravity theory, like those arising in the Palatini $f(\mathbf{R})$
theory, generally possesses a scale-dependent effective
sound-speed-squared term which affects the growth of CDM density
perturbations and thus influences the matter power spectrum
\cite{Koivisto2006, Li2006a, Li2006b} on small scales.

\section{The Cosmological Background Evolution}

\label{sect:Background}

In order to analyse the cosmological background evolution we can
neglect the $q_{a}^{f}$ and $\pi_{ab}^{f}$ terms, and hence the
quantities $\Upsilon_{a},\Sigma_{ab}$. As a result, the equations
are greatly simplified.

We are interested in the modified Friedmann equation in the
present model. From Eqs.~(\ref{eq:8}, \ref{eq:11}, \ref{eq:33}) we
have
\begin{eqnarray}  \label{eq:36}
3H^{2}&=& \kappa \rho ^{\mathrm{tot}},
\end{eqnarray}
in which $H\equiv \frac{1}{3}\theta $ is the Hubble expansion rate
and $\rho^{\mathrm{tot}}$ is expressed as
\begin{eqnarray}  \label{eq:37}
\kappa \rho ^{\mathrm{tot}} &=& \kappa
\rho^{f}+\frac{1}{2}f-2F\Delta ^{2}+ \frac{1}{2}\delta R-\delta
R_{ab}u^{a}u^{b},
\end{eqnarray}
with $\delta R$ and $\delta R_{ab}$ given in Eqs.~(\ref{eq:34},
\ref{eq:35}). After a lengthy calculation, and using
Eq.~(\ref{eq:9}) to eliminate the term
$\dot{\theta}+3\ddot{\lambda}/2\lambda $ which appears in
Eq.~(\ref{eq:37}), we obtain the following simple result
\begin{eqnarray}  \label{eq:38}
\left[ H+\frac{\dot{\lambda}}{2\lambda }\right]^{2} &=&
\frac{1}{6}(\Delta -3\omega \Xi ).
\end{eqnarray}
There are two interesting points regarding Eq.~(\ref{eq:38}).
Firstly, we see that only $\omega $, and not its time derivatives
$\dot{\omega}$ or $\ddot{\omega},$ enter the equation. Secondly,
the second-order derivative of $\lambda$ does not appear either;
to see the consequence of this, note that since
$\dot{\lambda}=\partial \lambda (\rho ^{f})/\partial \rho
^{f}\dot{\rho}^{f}=-\varsigma \partial \lambda (\rho
^{f})/\partial \rho ^{f}\rho ^{f}H$ (with $\varsigma =3$ for
matter and $\varsigma =4$ for radiation), the dependence of
$\lambda $ on $p^{f}$ could be expressed in terms of $\rho^{f}$,
\emph{e.g.}, in radiation-dominated era $p^{f}=\rho ^{f}$/3 and in
matter-dominated era $p^{f}=0$, and so we have
$\dot{\lambda}\propto H$. Consequently, Eq.~(\ref{eq:38}) has the
form
\begin{eqnarray}  \label{eq:39}
H^{2} &=& \Theta(\rho ^{f},p^{f}),
\end{eqnarray}
where $\Theta $ is a complicated function of $\rho ^{f}$ and
$p^{f}$ (at late times $p^{f}\doteq 0$ and it becomes a function
of $\rho ^{f}$ alone).

\begin{figure}[tbp]
\begin{center}
\includegraphics[scale=0.92] {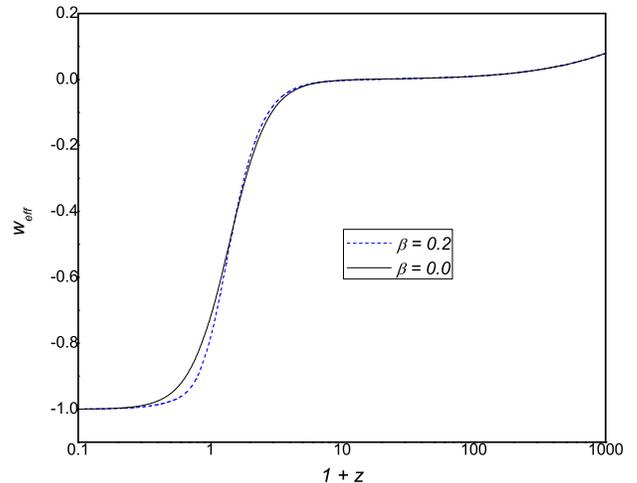}
\end{center}
\caption{(color online) The time evolution of the effective
equation of state for the $f(\mathbf{R}^{ab}\mathbf{R}_{ab}) =
\alpha(\mathbf{R}^{ab}\mathbf{R}_{ab})^{\beta}$ model. For
clearness we only show the cases of $\protect\beta =0.2$ (the
dashed curve) and $0$ ($\Lambda \mathrm{CDM}$, the solid curve).
The other parameters defining the cosmology are $\Omega _{m}=0.27$
and $\Omega _{r}=8.5\times 10^{-5}$. Clearly an accelerating phase
following the standard matter-dominated era can be realized within
this model. Note that the redshift range $1+z<1$ corresponds to
the future of the present epoch, where the Universe evolves into a
de Sitter stage.} \label{fig:Figure1}
\end{figure}

As we discussed above, knowing $\rho ^{f}$ and $p^{f}$ means that we know $%
\Delta ,\Xi ,\lambda $ and $\omega $. Thus, given a specific form for $f(%
\mathbf{R}^{ab}\mathbf{R}_{ab}),$ Eq.~(\ref{eq:38}) completely
determines the background cosmological evolution of the model. As
a particular example let us consider the case of
\begin{eqnarray}  \label{eq:40}
f(\mathbf{R}^{ab}\mathbf{R}_{ab}) &=& \alpha
(\mathbf{R}^{ab}\mathbf{R}_{ab})^{\beta }
\end{eqnarray}
where $\alpha $ and $\beta $ are the model parameters. Note that
here $\mathbf{R}^{ab}\mathbf{R}_{ab}=\Delta ^{2}+3\Xi ^{2}$ is
always non-negative, and $\beta =0$ corresponds to a picking the
standard $\Lambda \mathrm{CDM}$ cosmology of GR.

For convenience, we shall define the following dimensionless
quantities
\begin{eqnarray}  \label{eq:41}
\tilde{f} &\equiv &\frac{f}{H_{0}^{2}},  \nonumber \\
\tilde{\Delta} &\equiv &\frac{\Delta }{H_{0}^{2}},  \nonumber \\
\tilde{\Xi} &\equiv &\frac{\Xi }{H_{0}^{2}},  \nonumber \\
\tilde{F} &\equiv &FH_{0}^{2},  \nonumber \\
\Omega _{m} &\equiv &\frac{\kappa \rho _{m}}{3H_{0}^{2}},  \nonumber \\
\Omega _{r} &\equiv &\frac{\kappa \rho _{r}}{3H_{0}^{2}},  \nonumber \\
\tilde{\alpha} &\equiv &\alpha H_{0}^{4\beta -2},
\end{eqnarray}
then Eqs.~(\ref{eq:14}, \ref{eq:15}) could be rewritten as
\begin{eqnarray}  \label{eq:42}
\tilde{\Delta}+2\tilde{F}\tilde{\Delta}^{2}-\frac{1}{2}\left(\tilde{\Delta}
+3\tilde{\Xi}+\tilde{f}\right) &=& 3\Omega _{m}+3\Omega _{r}, \\
\label{eq:43}
\tilde{\Xi}+2\tilde{F}\tilde{\Xi}^{2}-\frac{1}{2}\left(
\tilde{\Delta} + 3\tilde{\Xi}+\tilde{f}\right) &=& -\Omega _{r},
\end{eqnarray}
where
\begin{eqnarray}  \label{eq:44}
\tilde{f}(\mathbf{R}^{ab}\mathbf{R}_{ab}) &=& \tilde{\alpha}\left(
\tilde{\Delta}^{2}+3\tilde{\Xi}^{2}\right) ^{\beta }, \\
\label{eq:45} \tilde{F}(\mathbf{R}^{ab}\mathbf{R}_{ab})
&=&\tilde{\alpha}\beta \left(
\tilde{\Delta}^{2}+3\tilde{\Xi}^{2}\right)^{\beta -1};
\end{eqnarray}
and than Eq.~(\ref{eq:38}) reduces to
\begin{eqnarray}  \label{eq:46}
\left[ 1-\frac{\varsigma
\lambda_{\rho^{f}}\rho^{f}}{2\lambda}\right]^{2}\frac{H^{2}}{H_{0}^{2}}=\frac{1}{6}\left(
\tilde{\Delta}-3\omega \tilde{\Xi}\right)
\end{eqnarray}
where $\lambda _{\rho ^{f}}\equiv \partial \lambda /\partial \rho
^{f}$.

In this paper we will set $\Omega _{r}=8.5\times 10^{-5}$, so
today we have $H^{2}/H_{0}^{2}=1$ and there are 3 equations
(Eqs.~(\ref{eq:42}, \ref{eq:43}, \ref{eq:46})) for the 5
parameters $\tilde{\Delta},\tilde{\Xi},\tilde{\alpha},\beta $ and
$\Omega _{m}$. Therefore, we are able to express all the other
quantities in terms of $\beta $ and $\Omega_{m}$, which can
therefore be treated as the two independent degrees of freedom of
our model. Note that $\tilde{\alpha}$ is a constant, and once
evaluated at the present day, it could be used all through the
cosmic history, which helps determine $\tilde{\Delta},\tilde{\Xi}$
at arbitrary times.

In Figure~\ref{fig:Figure1} we have plotted the effective equation
of state, defined by
$w_{eff}\equiv-1-\frac{2}{3}\frac{\dot{H}}{H^{2}} = -1-
\frac{2}{3}\frac{H^{\ast}}{H}$ (where a star-superscript denotes
the derivative with respect to $\log (a)$), as a function of the
redshift. The values of $\beta $ are indicated beside the curves.
At early times, when the $f(\mathbf{R}^{ab}\mathbf{R}_{ab})$
corrections are negligible, the models all mimic the $\Lambda
\mathrm{CDM}$ evolution of $w_{eff}$, and the same thing happens
in the future. This is because during this era the matter
(relativistic and non-relativistic) is greatly diluted so that the
right-hand sides of Eqs.~(\ref{eq:42}, \ref{eq:43}) both vanish;
consequently, we can solve them to show that
$\tilde{\Delta}=\tilde{\Xi}=\mathrm{const.}$ and so
$f(\mathbf{R}^{ab}\mathbf{R}_{ab})$ is also constant. The
deviation from $\Lambda \mathrm{CDM}$ occurs mainly at
intermediate times, that is, in the recent past and future.

We now use the observational data on the background cosmology to
constrain the parameter space (in the $\beta -\Omega _{m}$ plane)
of the present model. For this we jointly use the 157 measurements
on SNe luminosity distance in the Gold data sets of Riess
\emph{et~al}. \cite{Riess2004} and the CMB shift (R) parameter.
The SNe luminosity distance is expressed as
\begin{eqnarray}  \label{eq:47}
d_{L}(z) &=& (1+z)\int_{0}^{z}\frac{du}{H(u)}  \nonumber \\
&=&\frac{1+z}{H_{0}}\int_{0}^{z}\frac{du}{E(u)}
\end{eqnarray}
where $E(z)=H(z)/H_{0}$. The measurements supply the
extinction-corrected distance modulus $\mu _{0}=5\lg d_{L}+25$
(with $d_{L}$ in units of Mega-parsecs) and its uncertainty,
$\sigma$, for individual SNe, so that the standard $\chi ^{2}$
minimization, defined by
\begin{eqnarray}  \label{eq:48}
\chi ^{2} &=& \sum_{i=1}^{157}\frac{\left[
\mu_{p,i}(z_{i};H_{0},\Omega_{m},\beta )-\mu _{0,i}\right]
^{2}}{\sigma _{i}^{2}}
\end{eqnarray}
is easy to implement, where $\mu _{p}$ is the theoretically
predicted distance modulus. As $H_{0}$ appears only as it does in
Eq.~(\ref{eq:47}), we could marginalize over it by integrating the
probability density $p(\chi )\propto \exp (-\chi ^{2}/2)$ for all
values of $H_{0}$. For the CMB R-parameter, defined as
\begin{eqnarray}  \label{eq:49}
R &=& \sqrt{\Omega _{m}}H_{0}\int_{0}^{z_{dec}}\frac{dz}{H(z)},
\end{eqnarray}
we adopt the observational value $R^{\mathrm{obs}}=1.70\pm 0.03$
at $z_{dec}=1089$ from \cite{Wang2006}. Note that this does not
depend on the specified value of $H_{0}$.

\begin{figure}[tbp]
\centering
\includegraphics[scale=1.0] {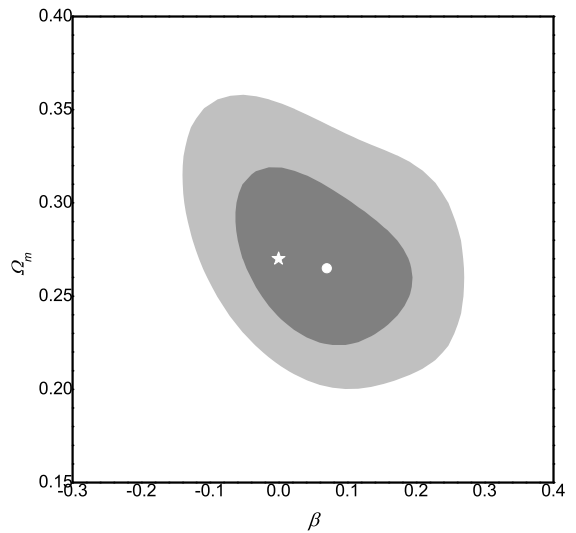}
\caption{The constraints on the parameter space of $\Omega_{m}$ and $\protect%
\beta$ in the present model from joint SNe and CMB shift parameter
data sets. The grey and light grey regions represent the 68.3\%
and 95.4\% confidence contours respectively. The white circle
($\Omega_{m} = 0.265, \protect\beta = 0.07$) is the best-fitting
parameter of our model, and the star is the concordance
$\Lambda\mathrm{CDM}$ model.} \label{fig:Figure2}
\end{figure}

Our constraining result is shown in Figure~\ref{fig:Figure2},
where we have shown the 68\% and 95\% confidence regions
respectively. The constrained intervals are roughly $0.20 \lesssim
\Omega_{m} \lesssim 0.36$ and $-0.13 \lesssim \beta \lesssim 0.29$
at the 95\% confidence level, with the best fitting values being
$(\beta, \Omega_{m}) \simeq (0.07, 0.265)$ with $\chi^2/dof \simeq
1.126$. Also note that the concordance $\Lambda\mathrm{CDM}$ model
(the white star) lies within the 68\% confidence region of our
constraints.

Thus, we see that the background cosmological data is able to
constrain $|\beta|$ to be of order $0.1$. In the next section we
will briefly investigate the possible constraint from the growth
of dark-matter density perturbations, and show that this may
provide a potentially more stringent restriction on $\beta$.
However, considering that this latter limit depends on the
properties of the dark matter, our background constraints given in
this section are less model dependent.

\section{Effects on Late-time CDM Density Perturbation Growth}

\label{sect:Perturbations}

In this section we study the effects of the
$f(\mathbf{R}^{ab}\mathbf{R}_{ab})$ corrections to GR on the CDM
density perturbation growth. We start by recalling the case of
$f(\mathbf{R})$ gravity because it shares some similarities with
the $f(\mathbf{R}^{ab}\mathbf{R}_{ab})$ one, while being
technically simpler than the latter, and because a similar
analysis for the former is still missing from Refs.~\cite{Li2006a,
Li2006b} (see however \cite{Koivisto2006} for a slightly different
treatment).

\subsection{The Case of Palatini $f(\mathbf{R})$ Gravity}

Recall that in our simplified model the universe is filled with
CDM and radiation, and at later times the radiation energy density
is negligible, so $\rho ^{f}\doteq \rho _{\mathrm{CDM}}$.

Taking the spatial derivative of the Raychaudhuri equation
Eq.~(\ref{eq:9}) (with the $\rho ,p$ there being replaced by
$\rho^{\mathrm{tot}}=\rho^{f}+\rho^{eff},p^{\mathrm{tot}}=p^{f}+p^{eff}$),
and working in the CDM frame (where the observer is comoving with
CDM particles and thus $A=0$) \cite{Lewis2000}, we have
\begin{eqnarray}  \label{eq:50}
\Delta _{\mathrm{CDM}}^{\prime \prime }+\mathcal{H}\Delta
_{\mathrm{CDM}}^{\prime }-\frac{\kappa
}{2}(\mathcal{X}^{\mathrm{tot}}+3\mathcal{X}^{p,
\mathrm{tot}})a^{2} &=& 0,
\end{eqnarray}
where $\Delta _{\mathrm{CDM}}$ is the CDM density perturbation
contrast that is defined through $\hat{\nabla}_{a}\rho
_{\mathrm{CDM}}=\rho _{\mathrm{CDM}}\sum_{k}\frac{k}{a}\Delta
Q_{a}^{k}$, and $\mathcal{H}=\frac{\theta }{3}a$ is the Hubble
expansion rate with respect to conformal time (note that a prime
denotes the conformal time derivative, and a dot the cosmic
comoving proper-time derivative); $\mathcal{X},\mathcal{X}^{p}$
are respectively the harmonic expansion coefficients for
$\hat{\nabla}_{a}\rho $ and $\hat{\nabla}_{a}p$ (defined via
$\hat{\nabla}_{a}\rho =\sum_{k}\frac{k}{a}\mathcal{X} Q_{a}^{k}$
and $\hat{\nabla}_{a}p=\sum_{k}\frac{k}{a}\mathcal{X}^{p}Q_{a}^{k}
$ \cite{Comment1}). Clearly we need to know about
$\mathcal{X}^{eff}$ and $\mathcal{X}^{p,eff}$ which arise from the
$f(\mathbf{R}^{ab}\mathbf{R}_{ab})$ modifications to GR
(c.f.~Eq.~(33)).

In the Palatini $f(\mathbf{R})$ model, in which the Ricci scalar
$\mathbf{R}$ in the gravitational action is replaced with
$\mathbf{R}+f(\mathbf{R})$, Eqs.~(\ref{eq:26}, \ref{eq:27}) still
hold, but with (see for example \cite{Li2006a})
\begin{eqnarray}  \label{eq:51}
\lambda &=& 1+\frac{\partial f}{\partial \mathbf{R}},  \nonumber \\
\xi _{ab} &=& 0,  \nonumber \\
\zeta _{ab} &=& 0.
\end{eqnarray}%
Then, with the help of the calculations in the appendix, it is
straightforward to show that
\begin{eqnarray}  \label{eq:52}
\mathbf{R}_{ab} &=&R_{ab}+\delta R_{ab}  \nonumber \\
&=& R_{ab}+\frac{3}{2\lambda ^{2}}\nabla _{a}\lambda
\nabla_{b}\lambda - \frac{1}{\lambda }\nabla _{a}\nabla_{b}\lambda
-\frac{1}{2\lambda}g_{ab}\Box \lambda \ \ \ \ \
\end{eqnarray}%
where $\Box = \nabla ^{2}$, and the modified Einstein equation
\cite{Li2006a},
\begin{eqnarray}
\lambda \mathbf{R}_{ab}-\frac{1}{2}g_{ab}(\mathbf{R}+f) &=& \kappa
T_{ab}^{f},  \nonumber
\end{eqnarray}
can be rewritten as
\begin{eqnarray}
R_{ab}-\frac{1}{2}g_{ab}R &=& \kappa T_{ab}^{\mathrm{tot}},
\nonumber
\end{eqnarray}
in which the effective total energy-momentum tensor is given by
\begin{eqnarray}  \label{eq:53}
&&\kappa T_{ab}^{\mathrm{tot}}  \nonumber \\
&=& \frac{1}{\lambda }\kappa T_{ab}^{f}+\frac{1}{2\lambda }g_{ab}(\mathbf{R}%
+f)-\frac{1}{2}g_{ab}(\mathbf{R}-\delta R)-\delta R_{ab}.\ \ \ \ \
\end{eqnarray}%
Using Eq.~(\ref{eq:8}), we can now identify
\begin{eqnarray}  \label{eq:54}
\kappa \rho ^{\mathrm{tot}} &=&\frac{1}{\lambda }\kappa \rho
^{f}+\frac{1}{2\lambda }(\mathbf{R}+f)  \nonumber \\
&&-\frac{1}{2}\left[ \mathbf{R}+\frac{3}{\lambda }\Box \lambda
-\frac{3}{2\lambda ^{2}}\nabla ^{a}\lambda \nabla _{a}\lambda \right]  \nonumber \\
&&-\frac{3\dot{\lambda}^{2}}{2\lambda^{2}}+\frac{\ddot{\lambda}}{\lambda }+%
\frac{1}{2\lambda }\Box \lambda, \\
\label{eq:55} \kappa p^{\mathrm{tot}} &=&\frac{1}{\lambda }\kappa
p^{f}-\frac{1}{2\lambda }(\mathbf{R}+f)  \nonumber \\
&&+\frac{1}{2}\left[ \mathbf{R}+\frac{3}{\lambda }\Box \lambda
-\frac{3}{2\lambda ^{2}}\nabla ^{a}\lambda \nabla _{a}\lambda \right]  \nonumber \\
&&-\frac{1}{3\lambda }(\theta
\dot{\lambda}+\hat{\nabla}^{2}\lambda)-\frac{1}{2\lambda }\Box
\lambda .
\end{eqnarray}%
Thus
\begin{eqnarray}  \label{eq:56}
\kappa
(\rho^{\mathrm{tot}}+3p^{\mathrm{tot}})=\frac{3}{\lambda}\ddot{\lambda}+\frac{1}{\lambda}\hat{\nabla}^{2}\lambda
+\cdots
\end{eqnarray}
in which $\cdots$ represent the terms \emph{not} involving second
order (time and spatial) derivatives.

The reason why we keep only two second derivative terms explicitly
on the right-hand side of Eq.~(\ref{eq:56}) is that, after taking
the spatial covariant derivative, the first term contributes a
$\Delta _{\mathrm{CDM}}^{\prime \prime }$ piece to
Eq.~(\ref{eq:50}) while the second term contributes a
$k^{2}\Delta_{\mathrm{CDM}}$ piece. None of the remaining terms in
$\cdots$ contribute these two pieces to Eq.~(\ref{eq:50}). To be
more explicit, recall that $\lambda =\lambda
(\rho_{\mathrm{CDM}})$ in the model, so
\begin{eqnarray}  \label{eq:57}
\hat{\nabla}_{a}\lambda &=&\frac{\partial \lambda (\rho
_{\mathrm{CDM}})}{\partial \rho
_{\mathrm{CDM}}}\hat{\nabla}_{a}\rho _{\mathrm{CDM}}\nonumber\\
&=&\frac{\dot{\lambda}}{\dot{\rho}_{\mathrm{CDM}}}\hat{\nabla}_{a}\rho_{\mathrm{CDM}}\nonumber\\
&=&-\frac{\dot{\lambda}}{3H\rho_{\mathrm{CDM}}}\hat{\nabla}_{a}\rho_{\mathrm{CDM}}\nonumber\\
&=&-\frac{\dot{\lambda}}{3H}\sum_{k}\frac{k}{a}\Delta_{\mathrm{CDM}}Q_{a}^{k},
\end{eqnarray}
where we have used Eq.~(\ref{eq:10}) to background order. As a
result,
\begin{eqnarray}
&&\kappa
(\mathcal{X}^{\mathrm{tot}}+\mathcal{X}^{p,\mathrm{tot}})a^{2}\nonumber \\
&=& -\frac{\dot{\lambda}}{\lambda H}\Delta _{\mathrm{CDM}}^{\prime \prime }-%
\frac{\dot{\lambda}}{3\lambda
H}k^{2}\Delta_{\mathrm{CDM}}+\cdots,\nonumber
\end{eqnarray}%
and Eq.~(\ref{eq:50}) can be recast into the form
\begin{eqnarray}
\left[ 1+\frac{\dot{\lambda}}{2\lambda H}\right] \Delta
_{\mathrm{CDM}}^{\prime \prime }+[\cdots ]\Delta _{\mathrm{CDM}}^{\prime }  \nonumber \\
+\left[ \cdots +\frac{\dot{\lambda}}{6\lambda H}k^{2}\right]
\Delta_{\mathrm{CDM}} &=& 0  \nonumber
\end{eqnarray}%
which, after rearrangement, gives
\begin{eqnarray}  \label{eq:58}
\Delta _{\mathrm{CDM}}^{\prime \prime }+[\cdots ]\Delta_{\mathrm{CDM}}^{\prime }  \nonumber \\
+\left[\cdots +\frac{\dot{\lambda}}{3(2\lambda
H+\dot{\lambda})}k^{2}\right] \Delta_{\mathrm{CDM}} &=& 0.
\end{eqnarray}%
Here, $\cdots$ denotes complicated terms that are determined
completely by the background evolutions of the model, and are
unimportant for our analysis here. What is essential in
Eq.~(\ref{eq:58}) is that it tells us that, as long as the
quantity $\dot{\lambda}$ does not vanish, in general there will
appear an effective sound-speed-squared term for the growth of
matter density perturbations. Depending on the sign of
$\dot{\lambda}$, this sound-speed-squared term could be either
positive or negative, in both cases the small-scale density
perturbation growth becomes extremely scale-dependent, altering
the shape of the matter power spectra significantly
\cite{Koivisto2006, Li2006a, Li2006b}. Notice that the terms in
$[\cdots]$ could also modify the evolution of density contrasts
differently as compared with the prediction in standard general
relativity, but in a scale-independent manner, and at small scales
their effects are subdominant.

One more comment on the modified Friedmann equation in the
Palatini gravity models is appropriate. Using Eq.~(\ref{eq:11})
with $\rho $ replaced by the $\rho ^{\mathrm{tot}}$ given in
Eq.~(\ref{eq:54}), we can see that only the $\dot{\lambda}\theta
,\dot{\lambda}^{2}$ terms are involved and the $\ddot{\lambda}$
terms cancel (we only consider terms to background order here).
Since $\dot{\lambda}=\partial \lambda (\rho ^{f})/\partial \rho
^{f}\dot{\rho}^{f}=-\varsigma \partial \lambda (\rho^{f})/\partial
\rho ^{f}\rho ^{f}H$ we see that $\dot{\lambda}\theta,
\dot{\lambda}^{2}\propto H^{2}$ and could be moved to the
left-hand side of Eq.~(11); the remaining terms on the right-hand
side are also functions of $\rho ^{f}$ only and so
Eq.~(\ref{eq:39}) is also realized. To be more explicit, the
modified Friedmann equation in Palatini $f(\mathbf{R})$ gravity is
\cite{Li2006a}
\begin{eqnarray}
\left[ H+\frac{\dot{\lambda}}{2\lambda }\right] ^{2} &=&
\frac{1}{6\lambda}\left[ \kappa
(\rho^{f}+3p^{f})-(\mathbf{R}+f)\right]  \nonumber
\end{eqnarray}
which can be shown to be just Eq.~(\ref{eq:38}) if $\omega =1$
there, as expected, because the metrics then take the same form
(of course, the definitions of $\lambda $ are different in the two
cases).

\subsection{The Case of Palatini $f(\mathbf{R}^{ab}\mathbf{R}_{ab})$ Gravity}

Now consider the $f(\mathbf{R}^{ab}\mathbf{R}_{ab})$ gravity
model. As discussed in the last section, the detailed forms of
$\rho ^{eff},p^{eff}$ are very complicated, but fortunately we
need not evaluate the full formulae explicitly. Our experience of
the simpler theory described in the last subsection shows that
what is most relevant for our analysis are the second order (time
and space) derivative terms (note that the term $\dot{\theta}$, if
exists, can also contribute to $\ddot{\Delta}_{\mathrm{CDM}}$
because $\hat{\nabla}_{a}\dot{\theta}$ contains
$\dot{\mathcal{Z}}$ where $\mathcal{Z}$ is the Harmonic expansion
coefficient of $\hat{\nabla}_{a}\theta$ via
$\hat{\nabla}_{a}\theta =
\frac{k^{2}}{a^{2}}\mathcal{Z}Q^{k}_{a}$, and because
$a\dot{\Delta}_{\mathrm{CDM}} = -k\mathcal{Z}$; however it turns
out no $\dot{\theta}$ terms appears in
$\kappa(\rho^{\mathrm{tot}}+3p^{\mathrm{tot}})$), which are
straightforward to identify.

We shall formally repeat the procedure of the last subsection.
Note that the quantities $\Upsilon _{a}, \Sigma _{ab}$ are
determined by $q_{a}^{f}, \pi _{ab}^{f}$ which in our case are due
to the radiation matter species. At late times the radiation
energy density is negligible so that, to a good approximation,
$q_{a}^{f}, \pi _{ab}^{f}$ and thus $\Upsilon _{a}, \Sigma _{ab}$,
vanish. As a result, Eqs.~(\ref{eq:26}, \ref{eq:27}) become
\begin{eqnarray}  \label{eq:59}
\mathbf{g}_{ab} &=& \lambda g_{ab}+\lambda (\omega -1)u_{a}u_{b}, \\
\label{eq:60} \mathbf{g}^{ab} &=& \frac{1}{\lambda
}g^{ab}+\frac{1}{\lambda }\left( \frac{1}{\omega }-1\right)
u^{a}u^{b},
\end{eqnarray}%
with $\lambda ,\omega $ defined in Eqs.~(\ref{eq:28},
\ref{eq:29}). Meanwhile, since $\rho^{f}\doteq
\rho_{\mathrm{CDM}},$ we have $p^{f}\doteq 0$ at late times; $F,
\Delta ,\Xi $ , and hence $\lambda $ and $\omega ,$ become
functions of $\rho _{\mathrm{CDM}}$ only, \emph{i.e.}, $\lambda
=\lambda (\rho _{\mathrm{CDM}})$, $\omega =\omega (\rho
_{\mathrm{CDM}})$. The analysis in Eq.~(\ref{eq:57}) then also
applies to $\lambda $ and $\omega $ here, so
\begin{eqnarray}
\hat{\nabla}_{a}\lambda &=&
-\frac{\dot{\lambda}}{3H}\sum_{k}\frac{k}{a}
\Delta_{\mathrm{CDM}}Q_{a}^{k},  \nonumber \\
\hat{\nabla}_{a}(\lambda \omega ) &=&
-\frac{(\lambda\omega)^{\cdot
}}{3H}\sum_{k}\frac{k}{a}\Delta_{\mathrm{CDM}}Q_{a}^{k}. \nonumber
\end{eqnarray}

Now, from Eq.~(\ref{eq:33}) we obtain
\begin{eqnarray}  \label{eq:61}
&&\kappa (\rho ^{\mathrm{tot}}+3p^{\mathrm{tot}})  \nonumber \\
&=& 2\kappa \rho ^{\mathrm{tot}}-\kappa(\rho
^{\mathrm{tot}}-3p^{\mathrm{tot}}) \nonumber \\
&=& \kappa (\rho ^{f}+3p^{f})-2F(\Delta ^{2}-3\Xi ^{2})-f-2\delta
R_{ab}u^{a}u^{b}\ \ \
\end{eqnarray}%
where the relation $\mathbf{R}_{a}^{c}\mathbf{R}_{cb}\doteq \Delta
^{2}u_{a}u_{b}+\Xi ^{2}h_{ab}$ and Eq.~(\ref{eq:8}) are used;
$\delta R_{ab}$ is given in Eq.~(\ref{eq:34}). Note that $\delta
R$ does not appear in this formula, and what we need to evaluate
is just the collection of second-order derivative terms in $\delta
R_{ab}u^{a}u^{b}$. After some manipulation we obtain a similar
expression to Eq.~(\ref{eq:56}) for the Palatini $f(\mathbf{R})$
model:
\begin{eqnarray}  \label{eq:62}
\kappa (\rho ^{\mathrm{tot}}+3p^{\mathrm{tot}})=\frac{3}{\lambda
}\ddot{\lambda}+\frac{1}{\lambda }\hat{\nabla}^{2}(\lambda \omega
)+\cdots
\end{eqnarray}
The following analysis then completely parallels that for the
Palatini $f(\mathbf{R})$ model and the CDM density perturbation
growth equation can be shown to be (like Eq.~(\ref{eq:58}))
\begin{eqnarray}  \label{eq:63}
\Delta _{\mathrm{CDM}}^{\prime \prime }+[\cdots ]\Delta _{\mathrm{CDM}}^{\prime }  \nonumber \\
+\left[\cdots +\frac{(\lambda \omega )^{\cdot }}{3(2\lambda
H+\dot{\lambda})}k^{2}\right] \Delta _{\mathrm{CDM}} &=& 0.
\end{eqnarray}
Again the $[$ $\cdots ]$ in Eq.~(\ref{eq:63}) denotes the terms
which are completely determined by the background evolution and
are not of interests to us here. Thus we see that, similar to the
case of Palatini $f(\mathbf{R})$ gravity model, a scale-dependent
sound-speed-squared term also appears in the Palatini
$f(\mathbf{R}^{ab}\mathbf{R}_{ab})$ gravity model, whose sign
depends on $(\lambda \omega )^{\cdot }$. Note that in the case of
$\omega =1$ the metric $\mathbf{g}_{ab}$ (Eq.~(\ref{eq:59},
\ref{eq:60})) has the same form as that in the Palatini
$f(\mathbf{R})$ model, and Eq.~(63) reduces to Eq.~(58) as
expected.

In the general $f(\mathbf{R}^{ab}\mathbf{R}_{ab})$ gravity models
it is possible that $(\lambda \omega )^{\cdot }\neq 0$, thus the
scale-dependent effective sound-speed-squared term influences the
matter perturbation growth and could alter the shape of the
predicted matter power spectrum. Our previous knowledge derived
from refs. \cite{Koivisto2006, Li2006a, Li2006b} suggests that
this effect might allow observational data to place stringent
constraints on the parameter space of these theories. In fact, we
can give a rough estimate of how stringent the constraint can be.
Consider first Eq.~(\ref{eq:58}) for the Palatini $f(\mathbf{R})$
model: since on small scales the $k^{2}$ term dominates the other
terms in front of $\Delta _{\mathrm{CDM}}$ (the
$^{\prime}\cdots^{\prime }$ terms), it is obvious that the
magnitude of the quantity $\dot{\lambda}/3(2\lambda
H+\dot{\lambda})$ determines the deviation from $\Lambda
\mathrm{CDM}$ results. The observational constraint on the model
parameter $\beta $ (recall that $f(\mathbf{R})=\alpha
(-\mathbf{R})^{\beta }$), as shown in Refs.~\cite{Koivisto2006,
Li2006a}, is $|\beta |<\mathcal{O}(10^{-6}\sim 10^{-5})$; we take
$|\beta|\sim 10^{-5}$ and $\Omega _{m}=0.3$ for illustrative
purposes, and find that $|\dot{\lambda}/3(2\lambda
H+\dot{\lambda})|\sim \mathcal{O}(10^{-7}-10^{-6})$ for the
relevant redshift range of $0\lesssim z\lesssim 10$. In the
analogous case of Palatini $f(\mathbf{R}^{ab}\mathbf{R}_{ab})$
gravity, Eq.~(\ref{eq:63}), we have $(\lambda \omega )^{\cdot
}/3(2\lambda H+\dot{\lambda})$ as the dominant term instead of
$\dot{\lambda}/3(2\lambda H+\dot{\lambda})$, and so the constraint
should be $|(\lambda \omega )^{\cdot}/3(2\lambda
H+\dot{\lambda})|<\mathcal{O}(10^{-7}-10^{-6})$ in the same
redshift range. Again, taking $\Omega _{m}=0.3$, for the theories
with $f(\mathbf{R}^{ab}\mathbf{R}_{ab})=\alpha
(\mathbf{R}^{ab}\mathbf{R}_{ab})^{\beta }$ we find that to satisfy
the above constraint $|\beta |$ must also be limited to be
$\mathcal{O}(10^{-5})$ or even smaller.

Thus, we conclude that the Palatini
$f(\mathbf{R}^{ab}\mathbf{R}_{ab})$ models may be constrained by
the observational data on the matter power spectrum just as
stringently as are the Palatini $f(\mathbf{R})$ models. Yet, we
should note that a more exact quantitative constraint can only be
obtained by exploiting a full parameter-space search as done in
\cite{Li2006a}, and that the above conclusion depends on the
assumption that the CDM particles have vanishing anisotropic
stress. If, in contrast, the dark matter particles admit an
anisotropic stress, in a manner similar to that prescribed in
\cite{Koivisto2007}, then the effective sound-speed-squared terms
might be canceled and leave no significant traces.

\section{Discussion and Conclusions}

\label{sect:Conclusions}

We considered a general class of modified gravity models where the
Ricci scalar in the gravitational action of GR is replaced by a
function $\mathbf{R}+f(\mathbf{R}^{ab}\mathbf{R}_{ab})$ and the
field equations are derived using the Palatini variational
approach, \emph{i.e.}, treating the metric $g_{ab}$ and connection
$\mathbf{\Gamma }_{bc}^{a}$ as independent variables so that the
action is varied with respect to both of them. The strategy for
deriving the cosmological equations at both the background
(zero-order) and the first-order perturbation levels is outlined.
The main step in this process is to determine the metric,
$\mathbf{g}_{ab}$, whose Levi-Civita connection is $\mathbf{\Gamma
}_{bc}^{a}$, relate it to the physical metric $g_{ab}$ and thereby
fix the relation between $\mathbf{R}_{ab}$ and $R_{ab}$. Then, the
correction to GR is treated as a new effective energy-momentum
tensor while the field equations take the same form as in GR. The
formulae laid down here might be useful for the numerical
implementations of such modified gravity models.

We also investigated in detail a power-law correction to the usual
Einstein-Hilbert action given by
$f(\mathbf{R}^{ab}\mathbf{R}_{ab})=\alpha
(\mathbf{R}^{ab}\mathbf{R}_{ab})^{\beta }$. We used the SNIa
luminosity distance and CMB shift parameter data to constrain its
(independent) two-parameter space $(\beta ,\Omega _{m})$. We found
that at 95\% confidence level $\beta \in \lbrack -0.13,0.29]$ and
$\Omega_{m}\in \lbrack 0.20,0.36]$. A slightly positive value of
$\beta $ ($\beta \simeq 0.07$) is preferred by the data used.
However, the standard $\Lambda \mathrm{CDM}$ model (equivalent to
$\beta =0$) with $\Omega_{m}=0.27$ is still within the 68\%
confidence contour. Hence, although the best fit to the data is a
non-null leading-order correction to Einstein gravity, the current
data exhibits no significant preference over the concordance
$\Lambda \mathrm{CDM}$ model of GR.

The late-time growth of matter density perturbations in general
Palatini $f( \mathbf{R}^{ab}\mathbf{R}_{ab})$ gravity models was
also studied. It was shown that the equations governing this class
of models look very similar to that in the Palatini
$f(\mathbf{R})$ models. In particular, there exists a
scale-dependent effective sound-speed-squared term in the
perturbation growth equation which may be either positive or
negative, depending on the background evolution in both models. In
the $f(\mathbf{R})$ case it is well known that these terms can
lead to strong scale-dependence of the matter power spectrum,
which is highly constrained by observational data \cite
{Koivisto2006, Li2006a}, and we expect a similar feature to exist
in the $f(\mathbf{R}^{ab}\mathbf{R}_{ab})$ case. We estimate that
this will produce a strong observational bound of $|\beta|
\lesssim \mathcal{O}(10^{-5})$ unless some exotic properties are
added to the dark matter candidate \cite{Koivisto2007}.

As the final remark, we give a brief comment about the static and
spherically-symmetric solutions of the present model. The analogue
for the Palatini $f(\mathbf{R})$ model was considered in
\cite{Barausse2007} and the authors found that the exterior
spherically-symmetric vacuum solutions are unique. Here we just
want to point out the $f(\mathbf{R}^{ab}\mathbf{R}_{ab}) $ model
also shares this feature. In fact, in the vacuum where
$\rho^{f}=p^{f}=0$ it is easy to show that for our model $\Delta
=\Xi =\mathrm{const}.$ are uniquely determined by
Eqs.~(\ref{eq:14}, \ref{eq:15}) and so is $\mathbf{R}_{ab}$. The
full consideration of a static system  also needs the interior
solution and its matching  the exterior, which is beyond the scope
of this paper and will be left for further investigation.

\section*{Appendix}

\label{sect:Appendix}

In this Appendix we present the relation between
$\mathbf{R}_{ab}(\mathbf{\Gamma}^{a}_{bc})$ and $R_{ab}(g_{ab})$
if the two metrics $\mathbf{g}_{ab}$ and $g_{ab} $ satisfy the
following relations
\begin{eqnarray}  \label{eq:A2}
\mathbf{g}_{ab} &=&\lambda g_{ab}+\xi _{ab}, \\
\mathbf{g}^{ab} &=&\frac{1}{\lambda }g^{ab}+\zeta ^{ab}
\end{eqnarray}%
where $\lambda $ is a scalar function and $\xi _{ab},\zeta _{ab}$
symmetric tensors.

Firstly, the requirement
\begin{eqnarray}  \label{eq:A3}
\mathbf{g}^{ac}\mathbf{g}_{cb} = g^{ac}g_{cb} = \delta _{b}^{a}
\end{eqnarray}
implies that
\begin{eqnarray}  \label{eq:A4}
\lambda \zeta _{ab} + \frac{1}{\lambda }\xi _{ab} +
\zeta_{a}^{c}\xi_{cb} &=& 0.
\end{eqnarray}
Then, with some algebra, and using Eq.~(\ref{eq:A4}), we can
easily show that (here a comma denotes the ordinary derivative)
\begin{eqnarray}  \label{eq:A5}
&&\mathbf{\Gamma }_{bc}^{a}(\mathbf{g}_{ab})  \nonumber \\
&=&
\frac{1}{2}\mathbf{g}^{ad}(\mathbf{g}_{bd,c}+\mathbf{g}_{cd,b}+\mathbf{g}_{bc,d})  \nonumber \\
&=& \frac{1}{2}\left( \frac{1}{\lambda }g^{ad}+\zeta ^{ad}\right)
\times
\nonumber \\
&& \left[ (\lambda g_{bd}+\xi _{bd})_{,c}+(\lambda g_{cd}+\xi_{cd})_{,b}+(\lambda g_{bc}+\xi _{bc})_{,d}\right]  \nonumber \\
&=&\Gamma _{bc}^{a}(g_{ab})+\gamma _{bc}^{a},
\end{eqnarray}%
where the difference between $\mathbf{\Gamma }_{bc}^{a}$ and
$\Gamma {bc}^{a}$, denoted $\gamma _{bc}^{a},$ is defined by
\begin{eqnarray}  \label{eq:A6}
\gamma _{bc}^{a} &\equiv &\frac{1}{2\lambda }\left[
\delta_{b}^{a}\nabla_{c}\lambda +\delta _{c}^{a}\nabla _{b}\lambda
-g_{bc}\nabla ^{a}\lambda\right]  \nonumber \\
&&+\frac{1}{2}\left[ \zeta _{b}^{a}\nabla _{c}\lambda
+\zeta_{c}^{a}\nabla_{b}\lambda -g_{bc}\zeta ^{ad}\nabla _{d}\lambda \right]  \nonumber \\
&&+\frac{1}{2\lambda }\left[ \nabla _{b}\xi _{c}^{a}+\nabla_{c}\xi_{b}^{a}-\nabla ^{a}\xi _{bc}\right]  \nonumber \\
&&+\frac{1}{2}\zeta^{ad}[\nabla_{c}\xi_{bd}+\nabla_{b}\xi_{cd}-\nabla_{d}\xi_{bc}]
\end{eqnarray}%
and is a true tensor, as expected. From the definition in
Eq.~(\ref{eq:2}) we can thus derive that
\begin{eqnarray}  \label{eq:A7}
\mathbf{R}_{ab}=R_{ab}+\nabla_{c}\gamma_{ab}^{c}-\nabla_{b}\gamma_{ac}^{c}+\gamma_{ab}^{d}\gamma_{cd}^{c}-\gamma_{ac}^{d}\gamma_{bd}^{c}
\end{eqnarray}
Note that $\mathbf{R}_{ab}$ differs from $R_{ab}$ by a rank-2
symmetric tensor.

\

\begin{acknowledgments}

We thank Andrzej Borowiec, John Miritzis, Thomas Sotiriou and the
referee for helpful discussions. BL is supported by the Overseas
Research Studentship, Cambridge Overseas Trust and the Department
of Applied Mathematics and Theoretical Physics (DAMTP) at the
University of Cambridge. DFM acknowledges the A. Humboldt
Foundation.
\end{acknowledgments}

\bigskip



\begin{thebibliography}
\bibitem[Commet1(2007)]{Comment1} Here $Q^{k}_{a} = \frac{a}{k}\hat{\nabla}_{a}Q^{k}$ and $Q^{k}$ is the zero order eigenfunction of the comoving
Laplacian $a^{2}\hat{\nabla}^{2}$ ($a^{2}\hat{\nabla}^{2}Q^{k} =
k^{2}Q^{k}$), with eigenvalue $k^{2}$, $k$ representing a
specified scale.
\end{thebibliography}
\end{document}